\documentclass[sigconf]{acmart}
\AtBeginDocument{%
  }

\copyrightyear{2025}
\acmYear{2025}
\setcopyright{acmlicensed}\acmConference[ICAIF '25]{6th ACM International Conference on AI in Finance}{November 15--18, 2025}{Singapore, Singapore}
\acmBooktitle{6th ACM International Conference on AI in Finance (ICAIF '25), November 15--18, 2025, Singapore, Singapore}
\acmDOI{10.1145/3768292.3770421}
\acmISBN{979-8-4007-2220-2/2025/11}

\usepackage[utf8]{inputenc} 
\usepackage[T1]{fontenc}    
\usepackage{hyperref}       
\usepackage{url}            
\usepackage{booktabs}       
\usepackage{amsfonts}       
\usepackage{nicefrac}       
\usepackage{microtype}      
\usepackage{xcolor}         
\usepackage{xcolor}
\definecolor{firstplace}{RGB}{139,0,0}    
\definecolor{secondplace}{RGB}{0,0,139}   
\definecolor{thirdplace}{RGB}{0,100,0}   
\usepackage{amsmath} 
\usepackage{multirow}
\usepackage{graphicx}
\usepackage{wrapfig}
\begin{document}

\title{DeltaLag: Learning Dynamic Lead-Lag Patterns in Financial Markets}

\author{Wanyun Zhou}
\affiliation{%
  \institution{The Hong Kong University of Science and Technology
 (Guangzhou)}
  \city{Guangzhou}
  \country{China}
}
\email{wzhou266@connect.hkust-gz.edu.cn}

\author{Saizhuo Wang}
\affiliation{%
  \institution{The Hong Kong University of Science and Technology}
  \city{Hong Kong SAR}
  \country{China}
}
\affiliation{%
  \institution{International Digital Economy Academy}
  \city{Shenzhen}
  \country{China}
}
\email{swangeh@connect.ust.hk}

\author{Mihai Cucuringu}
\affiliation{%
  \institution{University of California, Los
 Angeles}
  \city{Los
 Angeles}
  \country{USA}
}
\affiliation{
\institution{
Department of Statistics \& Oxford Man Institute of Quantitative Finance, 
University of Oxford}
\city{Oxford}
\country{UK}
}
\email{mihai@math.ucla.edu}

\author{Zihao Zhang}
\affiliation{%
  \institution{University of Oxford}
  \city{Oxford}
  \country{UK}
}
\email{zhangzihao@hotmail.co.uk}

\author{Xiang Li}
\affiliation{%
  \institution{The Hong Kong University of Science and Technology
 (Guangzhou)}
  \city{Guangzhou}
  \country{China}
}
\email{xli906@connect.hkust-gz.edu.cn}

\author{Jian Guo}
\affiliation{%
  \institution{International Digital Economy
 Academy}
  \city{Shenzhen}
  \country{China}}
\email{guojian@idea.edu.cn}

\author{Chao Zhang}
\authornote{Corresponding author.}
\affiliation{%
  \institution{The Hong Kong University of Science and Technology
 (Guangzhou)}
  \city{Guangzhou}
  \country{China}
}
\email{chaoz@hkust-gz.edu.cn}

\author{Xiaowen Chu}
\authornotemark[1]
\affiliation{%
  \institution{The Hong Kong University of Science and Technology
 (Guangzhou)}
  \city{Guangzhou}
  \country{China}
}
\email{xwchu@hkust-gz.edu.cn}

\begin{abstract}
The lead-lag effect, where the price movement of one asset systematically precedes that of another, has been widely observed in financial markets and conveys valuable predictive signals for trading. However, traditional lead-lag detection methods are limited by their reliance on statistical analysis methods and by the assumption of persistent lead-lag patterns, which are often invalid in dynamic market conditions. In this paper, we propose \textbf{DeltaLag}, the first end-to-end deep learning method that discovers and exploits dynamic lead-lag structures with pair-specific lag values in financial markets for portfolio construction. Specifically, DeltaLag employs a sparsified cross-attention mechanism to identify relevant lead-lag pairs. These lead-lag signals are then leveraged to extract lag-aligned raw features from the leading stocks for predicting the lagger stock's future return. Empirical evaluations show that DeltaLag substantially outperforms both fixed-lag and self-lead-lag baselines. In addition, its adaptive mechanism for identifying lead-lag relationships consistently surpasses precomputed lead-lag graphs based on statistical methods. Furthermore, DeltaLag outperforms a wide range of temporal and spatio-temporal deep learning models designed for stock prediction or time series forecasting, offering both better trading performance and enhanced interpretability.
\end{abstract}




\maketitle

\section{Introduction}
The study of lead-lag relationships in multivariate time series systems has attracted significant attention across various domains, including earth sciences \cite{DeLucaGiovanni,WuDi}, biology \cite{RungeJakob}, and especially financial markets \cite{buccheri2021high,albers2021fragmentation,tolikas2018lead,BennettStefanos,ZhangYichiDynamic,ZhangYichiRobust,qiLeadLag}. In particular, for stock markets, the lead-lag effect, where the price movement of one stock systematically precedes that of another, has been shown to be both statistically significant and economically meaningful  \cite{LiYongli,ShiDanni,MioriDeborah, fan_does_2022}. For instance, price changes in large-cap stocks such as Apple or Microsoft often precede those in smaller, sector-related stocks. This lead-lag effect may arise from underlying differences in liquidity, information flow, and investor attention. Harnessing these dynamic patterns thus offers a valuable source of predictive signals for quantitative investment.

Existing research on lead-lag-based trading strategies typically follows a two-step approach \cite{BennettStefanos}: first identifying lead-lag pairs using statistical methods, and second constructing trading signals for laggers based on the features of the identified leaders. These statistical methods are often favored in practice due to their interpretability and ability to produce signals with clear financial intuition compared to various deep learning models, which often function as `black boxes’ with limited economic rationale. However, they face two fundamental limitations. First, they rely on linear statistical measures, such as cross-correlation, which are often insufficient to identify lead-lag relationships, failing to capture the time-varying, nonlinear, and structurally complex interactions in real-world financial markets. Second, they assume temporal stability in lead-lag relationships, overlooking the fact that such dependencies are often transient and regime-dependent. This assumption fails to account for the weak momentum in asset correlations, where two stocks exhibiting strong correlation at a certain lag in the past are unlikely to maintain this relationship in the future \cite{ZhuHaoren,ZhengXiaolin}. As a result, strategies based on this rigid two-step process may not adapt well to dynamic markets and tend to deteriorate over time.

To address these challenges, we propose \textbf{DeltaLag}, the first deep learning method designed to adaptively and dynamically discover the cross-asset lead-lag relationship and leverage these lead-lag structures to construct the trading signal. DeltaLag generalizes the traditional notion of lead-lag by allowing each stock pair to be associated with a time-varying, pair-specific lag value, which we model as a learnable time \textbf{delta} representing the time offset between leader and lagger stocks. This formulation enables the model to capture asynchronous dependencies that evolve across both asset pairs and time. 
DeltaLag differs significantly from existing lead-lag detection models based on statistical analysis. These statistical methods first measure pairwise lead-lag metrics using different cross-correlation functions between stock return time series and then apply network clustering algorithms to identify leading and lagging clusters to form the final lead-lag graph. DeltaLag, instead, uses neural networks to adaptively capture asset interactions and construct a dynamic lead-lag graph on a daily basis. To achieve this, the model first applies temporal embedding to stock sequences, followed by a sparsified cross-attention mechanism that adaptively scores and selects candidate lead-lag pairs and their lag values.

Once the lead-lag pairs and their lag values are identified, we apply simple multilayer perceptron (MLP) networks to the leader stocks' raw price-volume features at their identified lag times to predict the lagger stock's future return ratio. This prediction step is similar to that in statistical lead-lag models, but differs fundamentally from many temporal and spatio-temporal deep learning models designed for stock return prediction or time series forecasting. Such deep learning models employ powerful sequence models like the encoder in Transformer to perform complex `black-box' operations on a stock's historical time series across multiple timesteps for its future return prediction. In contrast, lead-lag-based models only use the raw features of leader stocks from a specific day without leveraging the sequential information from the stock's own historical data, which provides a clear reflection of the underlying cross-asset lead-lag effect and thus offers greater interpretability. Despite the similarities between DeltaLag's prediction stage and that of statistical methods, a fundamental limitation of statistical lead-lag methods is that their detection and prediction stages are performed separately, preventing the detection process from being optimized toward the ultimate investment objective of achieving superior portfolio performance. Conversely, DeltaLag unifies these two stages into an end-to-end learning framework, allowing for the joint optimization of its detection and prediction modules during the neural network training process.
To enhance robustness in realistic trading scenarios, we further employ a ranking-based loss for cross-sectional stock selection, focusing on learning predicted return ordering to improve trading performance in our long-short portfolio. Our proposed architecture with this loss enhances modeling flexibility and enables the learned lead-lag relationships to be directly optimized for better portfolio performance.

Empirical results demonstrate that the lead-lag relationships identified by DeltaLag significantly enhance trading performance across multiple aspects.
Firstly, we show that allowing each stock pair to have a dynamically learned lag value leads to substantial improvements compared to using a fixed lag. Secondly, in line with some other lead-lag models \cite{BennettStefanos,ZhangYichiDynamic,MioriDeborah}, DeltaLag exclusively models cross-asset dependencies without relying on the target stock’s own historical data during the prediction stage. This design consistently outperforms self-lead-lag baselines, underscoring the crucial role of inter-stock dependencies in financial markets. Thirdly, when compared to a range of methods that use precomputed lead-lag graphs based on statistical analyses, DeltaLag with its adaptive property demonstrates superior performance. This highlights the limitations of existing lead-lag models that rely solely on correlation-based structures derived from historical data and confirms the weak momentum property of lead-lag relationships in financial markets. Finally, to further validate the robustness of our method, we compare it against a range of temporal and spatio-temporal baselines designed for stock prediction and time series forecasting. The consistent improvements in portfolio performance via backtesting validate the practical value of our adaptive, deep-learning-based lead-lag approach.
In summary, our contributions are as follows:
\begin{itemize}
     \item To the best of our knowledge, we are the first to propose an end-to-end deep learning framework that directly discovers time-varying, pair-specific lead-lag relationships in financial markets, addressing the limitations of traditional statistical analysis methods in capturing dynamic lead-lag structures.
     
    \item We propose a novel and end-to-end neural network 
    architecture that learns from price-volume features and uses cross-attention mechanisms to adaptively identify daily lead-lag pairs and their corresponding lag values, along with a specialized loss function tailored for cross-sectional stock selection.
    \item We empirically validate the strength of DeltaLag’s lead-lag detection by showing that it consistently outperforms fixed-lag and self-lead-lag baselines. Furthermore, we confirm the weak momentum of lead-lag relationships in financial markets by showing that our adaptively learned lead-lag graphs yield greater economic benefits than precomputed graphs derived from historical statistical correlations.
    \item Through extensive experiments, we demonstrate that DeltaLag surpasses a wide range of state-of-the-art temporal and spatio-temporal deep learning models for stock prediction or time series forecasting, offering both better trading performance and enhanced interpretability. 
    
\end{itemize}

\begin{figure*}[t]
    \centering
    \includegraphics[width=0.92\textwidth]{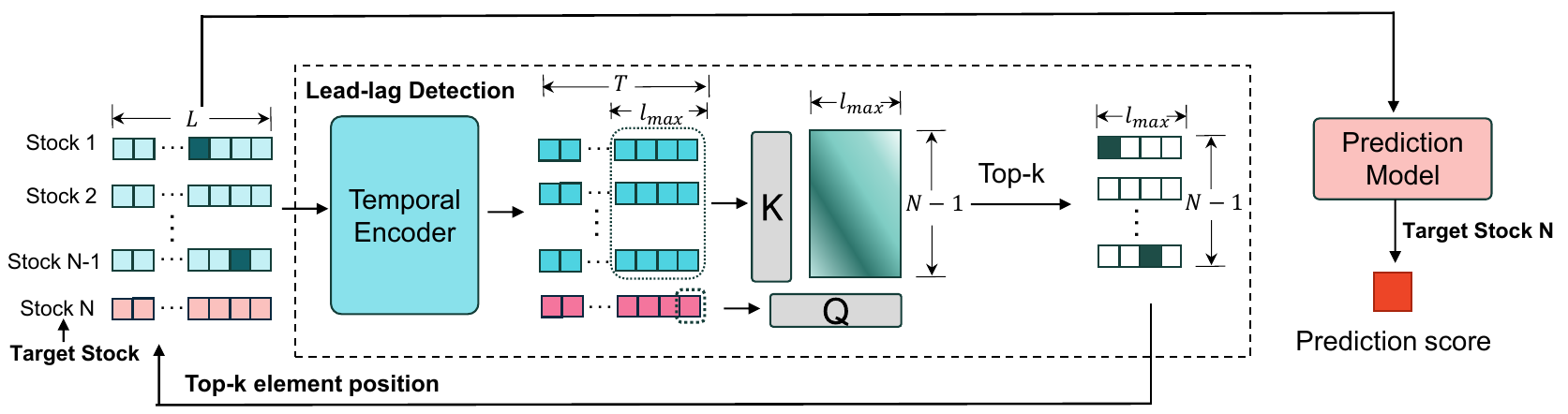}
    \caption{Architecture overview}
    \label{fig:arch}
\end{figure*}

\section{Related Work}
The study of lead-lag relationships among financial assets is a fundamental area of quantitative finance. Existing literature has approached this topic primarily through statistical methods, clustering techniques, and network analysis. 
Bennett et al. \cite{BennettStefanos} proposed a framework for detecting lead-lag structures by computing distance correlation and CCF-AUC between all stock pairs over lags from –5 to +5, followed by Hermitian clustering \cite{HermitianClust} to identify lead-lag clusters. They constructed a directed meta-flow graph based on inter-cluster flow imbalances, which exhibited strong predictive power in downstream financial forecasting tasks. Similarly, Li et al.\cite{LiYongli} analyzed the dynamic patterns of daily lead-lag networks specifically within the Chinese stock markets. Through statistical analysis, they discovered that successive lead-lag days follow a power-law distribution and provided a formal definition of the lead-lag effect through statistical testing. Their work employed exponential random graph models to identify factors influencing these lead-lag relationships. Another relevant line of research applies advanced temporal alignment techniques to enhance the robustness of lead-lag detection. Zhang et al. \cite{ZhangYichiDynamic} proposed utilizing Dynamic Time Warping (DTW), a powerful algorithm capable of aligning multivariate time series exhibiting temporal shifts and distortions. Their approach demonstrated robust lead-lag detection in lagged multi-factor models. Shi et al. \cite{ShiDanni}introduced a novel methodology based on multireference alignment, designed explicitly to address low signal-to-noise ratios scenarios. They also devise a cross-sectional trading strategy that capitalizes on lead-lag relationships. Moreover, the structural and dynamic complexity of lead-lag relationships has been explored by integrating macroeconomic regimes. Miori et al.\cite{MioriDeborah} studied the returns-driven macroeconomic regimes and their characteristic lead-lag behaviors among multiple asset classes. 

Despite these developments, traditional statistical and linear approaches remain susceptible to market non-stationarity and noise, motivating the need for adaptive, nonlinear, and deep-learning-based frameworks. However, very few studies have attempted to leverage deep learning for financial asset prediction by incorporating lead-lag properties, and those that do often approach the problem indirectly. 
Rooted in the broader causal discovery domain, Nauta et al. \cite{nauta2019causal} is most closely related to our work. They used convolutional neural networks for prediction and then attempt to infer temporal delays post-hoc by interpreting the model's internal parameters, specifically the convolutional kernel weights. Moving beyond this causal discovery perspective, Li et al. \cite{li2024stock} first used DTW algorithms to compute the lead-lag graph and subsequently apply graph neural networks to this pre-defined structure for prediction. Although this framework utilizes deep learning, the lead-lag detection stage itself does not. Therefore, our study aims to address these limitations, integrating recent advances in deep learning and adaptive cross-attention mechanisms to dynamically and directly uncover the lead-lag relationships and construct the trading signals.

\section{Problem Formulation}
Given a set of stocks $\mathcal{S}$, for each stock $u\in\mathcal{S}$ on trading day $t$, we collect price-volume features to form a feature vector $x_{u,t} \in \mathbb{R}^F$. Our objective is to predict the next-day return ratio of stocks.

For each target stock $u$ (lagger) \footnote{The `target stock' is the stock being predicted, also termed 'lagger' or 'follower' in lead-lag analysis; these terms are used interchangeably in our paper.} on day $t$, we aim to identify its top-$k$ leading stocks, denoted as $\mathcal{L}_{u,t} = \{v_1, v_2, \ldots, v_k\}$, along with their corresponding optimal lag values $\{\tau_{v_1 \to u}
, \tau_{v_2\to u}, \ldots, \tau_{v_k\to u}\}$. These top-$k$ leaders are defined as the stocks that exhibit the strongest lead-lag effects on the target stock $u$ at specific time lags.
Using these top-$k$ lead-lag pairs, we predict the return ratio of the lagger stock $u$ as:
\begin{equation}
\hat{r}_{u,t+1} = \mathcal{F}(\{x_{v,t-\tau_{v\to u}} | v \in \mathcal{L}_{u,t}\})
\end{equation}
where $\mathcal{F}$ is a prediction function (we use MLP in this paper) that maps the feature vectors of the leading stocks at their respective lagged time points to the predicted return ratio of the lagger stock.
This formulation captures the essence of our approach: using only information from other stocks (leaders) to predict the future returns of a target stock (lagger), without relying on the target stock's own historical data.

\section{Method}
\subsection{Lead-lag Detection Model}
To identify the top-$k$ leading stocks along with their corresponding lag values for a target stock $u$ on day $t$, we propose a self-adaptive cross-attention based lead-lag detection model. An overview of the DeltaLag architecture is shown in Figure \ref{fig:arch}. The detailed process is described as follows:

The input for our model consists of features derived from split-adjusted price and volume data, including several intraday price ratios (open, high, and low relative to the close), the daily return, log-transformed volume, and the daily turnover rate. Based on these features, for each stock $u \in \mathcal{S}$, we construct an input feature matrix $X_{u,t}$ over a rolling window of length $L$ from day $t-L+1$ to day $t$:
\begin{equation}
X_{u,t} = [x_{u,t-L+1}, x_{u,t-L+1}, \dots, x_{u,t}]^\top \in \mathbb{R}^{L \times F},
\end{equation}
where $F$ is the number of features. To extract meaningful temporal patterns from these sequences, we employ a temporal encoder $f_\theta$ (typically instantiated as sequence model such as LSTM) to transform the raw features $X_{u,t}$ into a hidden representation:
\begin{equation}
X'_{u,t} = f_\theta(X_{u,t}) \in \mathbb{R}^{L \times N},
\end{equation}
where $N$ is the hidden dimension of the temporal encoder. 

We then extract the embedding of the last timestep to form the query vector for stock $u$.
Let $X'_{u,t}[i,:] \in \mathbb{R}^{N}$ denote the embedding vector at the $i$-th timestep, with $i$ starting from 0 such that the first timestep corresponds to $i = 0$. We extract the last timestep embedding vector as follows:
\begin{equation}
h_{u,t}^{(L)} = X'_{u,t}[L-1,:] \in \mathbb{R}^{1 \times N}.
\end{equation}
We then compute the query vector for stock $u$ by applying a learnable weight matrix $W^Q \in \mathbb{R}^{N \times N}$:
\begin{equation}
q_{u,t} = (h_{u,t}^{(L)}W^Q)^T  \in \mathbb{R}^{1 \times N}.
\end{equation}

For all other stocks $v \in \mathcal{S} \setminus \{u\}$, we extract the embeddings corresponding to the last $l_{\text{max}}$ timesteps and transform them into key matrices:
\begin{equation}
\mathbf{K}_{v,t} = \mathbf{X}'_{v,t}[L-l_{\text{max}}:L,:] \mathbf{W}^K \in \mathbb{R}^{l_{\text{max}} \times N}, 
\end{equation}
where $\mathbf{W}^K \in \mathbb{R}^{N \times N}$ is another learnable weight matrix and $l_{\text{max}}$ represents the maximum lag value.
Notably, $l_{max}$ is usually smaller than $L$. A longer sequence $L$ provides sufficient context for temporal encoding, while restricting attention to the last $l_{max}$  steps focuses the model on the most likely lag candidates and improves both interpretability and efficiency. 

Then the attention scores between the target stock $u$ and each potential leading stock $v$ across different lag values can be computed as:
\begin{equation}
\mathbf{A}_{u,v,t} = \mathbf{q}_{u,t} \mathbf{K}_{v,t}^T \in \mathbb{R}^{1 \times l_{\text{max}}}.
\end{equation}
We then stack these attention scores for all potential leading stocks into a single attention score matrix:
\begin{equation}
\mathbf{A}_{u,t} = \begin{bmatrix}
\mathbf{A}_{u,v_1,t} \\
\mathbf{A}_{u,v_2,t} \\
\vdots \\
\mathbf{A}_{u,v_{|\mathcal{S}|-1},t}
\end{bmatrix} \in \mathbb{R}^{(|\mathcal{S}|-1) \times l_{\text{max}}}, 
\end{equation}
where $v_1, v_2, \ldots, v_{|\mathcal{S}|-1}$ represent all stocks in $\mathcal{S} \setminus \{u\}$.

From the attention score matrix $\mathbf{A}_{u,t}$, we identify the top-$k$ highest scores and their corresponding positions:
\begin{equation}
    \{(i_1, j_1), (i_2, j_2), \ldots, (i_k, j_k)\} = \text{TopK}(\mathbf{A}_{u,t}, k), 
\end{equation}
where each pair $(i_m, j_m)$ (for $m = 1, 2, \ldots, k$) indicates that stock $v_{i_m}$ is a leading stock for $u$ with a lag value of $\tau_{v_{i_m}\to u} = l_{\text{max}} - j_m$.
This gives us the set of top-$k$ leading stocks for target stock $u$:
\begin{equation}
    \mathcal{L}_{u,t} = \{v_{i_1}, v_{i_2}, \ldots, v_{i_k}\}, 
\end{equation}
with their corresponding optimal lag values:
\begin{equation}
    \{\tau_{v_{i_1}\to u}, \tau_{v_{i_2}\to u}, \ldots, \tau_{v_{i_k}\to u}\} = \{l_{\text{max}} - j_1, l_{\text{max}} - j_2, \ldots, l_{\text{max}} - j_k\}.
\end{equation}
Through this adaptive cross-attention mechanism, our model effectively and dynamically identifies lead-lag pairs and their lag values on each trading day, without relying on the precomputed lead-lag correlation graph.

\subsection{Signal Construction and Prediction}
After identifying lead-lag relationships, we construct trading signals to predict the return ratio for each target stock by using its detected leaders and their corresponding lag values.
For each target stock $u$ on day $t$, we have identified its top-$k$ leading stocks $\mathcal{L}_{u,t} = \{v_{i_1}, v_{i_2}, \ldots, v_{i_k}\}$ with their corresponding optimal lag values $\{\tau_{v_{i_1}\to u}, \tau_{v_{i_2}\to u}, \ldots, \tau_{v_{i_k}\to u}\}$ and attention scores:
\begin{equation}
    s_{v_{i_m},u} = \mathbf{A}_{u,t}[i_m, l_{\text{max}} - \tau_{v_{i_m}\to u}], 
\end{equation}
where $\mathbf{A}_{u,t}[i_m, l_{\text{max}} - \tau_{v_{i_m}\to u}]$ represents the element at position $(i_m, l_{\text{max}} - \tau_{v_{i_m}\to u})$ in the attention score matrix $\mathbf{A}_{u,t}$.
Then, we extract the raw price-volume features of these leading stocks at their respective lagged time points:
\begin{equation}
\mathbf{z}_{v_{i_m},u,t} = \mathbf{x}_{v_{i_m},t-\tau_{v_{i_m}\to u}} \in \mathbb{R}^F, 
\end{equation}
where $\mathbf{x}_{v_{i_m},t-\tau_{v_{i_m}\to u}}$ represents the feature vector of leading stock $v_{i_m}$ at time ($t-\tau_{v_{i_m}\to u}$).

To combine information from multiple leading stocks, we compute an attention-weighted sum of their features:
\begin{equation}
\mathbf{z}_{u,t} = \sum_{m=1}^{k} \frac{\exp(s_{v_{i_m},u})}{\sum_{m=1}^{k} \exp(s_{v_{i_m},u})} \mathbf{z}_{v_{i_m},u,t} \in \mathbb{R}^F,
\end{equation}
where the weights are derived from the softmax normalization of their attention scores. This ensures that leading stocks with stronger lead-lag relationships have a greater influence on the prediction.

The aggregated features are then fed into a multi-layer perceptron (MLP) to predict the next-day return ratio of the target stock:
\begin{equation}
\hat{r}_{u,t+1} = \text{MLP}(\mathbf{z}_{u,t})
\end{equation}
where the input dimension of MLP is $F$.

\subsection{Loss function design}
Since our primary objective is cross-sectional stock selection rather than predicting the exact magnitude of individual stock returns, we focus on correctly ranking stocks based on their expected returns. In this context, the \textit{relative rankings} of predicted returns is more important than their absolute values.
A natural approach would be to use a pairwise ranking loss that directly optimizes for correct ordering between pairs of stocks \cite{feng2019temporal}. This can be formulated as:
\begin{equation}
{L}_{\text{Pairwise}} = \sum_{i=1}^{N} \sum_{j=1}^{N} \max(0, -(\hat{r}_i - \hat{r}_j)(r_i - r_j)),
\end{equation}
where $\hat{r}_i$ is the network's predicted return score for stock $i$, $r_i$ is the actual return for stock $i$, and $N$ is the number of stocks. This loss function penalizes stock pairs where the predicted return scores and the actual returns have conflicting relative orderings, with penalties proportional to the magnitude of the mismatch.
However, recent research \cite{zhong_dspo_2024,WangSaizhuo} has demonstrated that such pairwise ranking losses can be challenging to optimize and may introduce numerical instability, further complicating the training process.
To address these issues, we adopt a more robust formulation based on a log-sum-exp pairwise structure. Theoretical analysis from the perspective of Bayes consistency has shown that this structure enhances optimization and improves numerical stability \cite{li2017improving}, thereby better achieving the ranking objective. 
Moreover, following \cite{zhong_dspo_2024}, we use the hyperbolic tangent function (tanh) as a smoother, differentiable transformation to capture the relative ordering relationships, which leads to the monotonic logistic regression loss:
\begin{equation}
{L}_{\text{Mon}} = \sum_{i=1}^{N} \sum_{j=1}^{N} \log (1 + \exp (-\tanh(\hat{r}_i - \hat{r}_j) \cdot \tanh(r_i - r_j))), 
\end{equation}
where $\hat{r}_i$ and $r_i$ represent the predicted and actual returns for stock $i$, respectively. This formulation encourages the model to learn monotonic relationships between predicted and actual returns while providing smoother gradients for more effective optimization and improve training stability.

\section{Experiments}

\subsection{Dataset and Experimental Setting}
We conducted our experiments using the dataset of US equities spanning from 2010 to 2023. Specifically, we used data from 2010 to 2018 as our training set, which included all available US equities during this period. The validation set comprised data from 2019 to 2021, while our test sets consisted of three different datasets from 2022 to 2023. We evaluated our method on stocks from the S\&P 500, NASDAQ market, and NYSE market. For the NASDAQ and NYSE datasets, we excluded stocks with market capitalization below 2B to ensure sufficient liquidity and trading volume for reliable analysis, resulting in universes of 713 and 1140 stocks, respectively. This multi-market evaluation allows us to verify the generalization ability of our method. 
In our implementation, we set $k=2$ for the top-$k$ leaders of the target stock. Setting $k$ greater than 1 avoids the non-differentiability issue inherent in top-$k$ positional selection, as it allows gradients to flow through the attention score matrix to the detection stage. We chose $k=2$ primarily for its computational efficiency, as complexity scales linearly with $k$. This decision was further supported by experiments confirming that the overall portfolio performance is nearly identical across larger values of $k$.\footnote{For $k$=2, 5, and 10, the average annualized returns across three datasets were 0.270, 0.273, and 0.262, with average Sharpe ratios of 2.53, 2.40, and 2.38, respectively.} 
Regarding the implementation, all baselines were trained and tuned on the same training and validation data splits as DeltaLag, and experiments were conducted on NVIDIA GeForce RTX 4090 GPUs, with data pre-processing and filtering following the QuantBench framework \cite{WangSaizhuo}. 

\subsection{Metrics}
To evaluate the performance of our model, we adopt three widely used metrics: Information Coefficient (\textbf{IC}), Annualized Return (\textbf{AR}), and Sharpe Ratio (\textbf{SR}). The IC measures the rank correlation between predicted and actual returns, reflecting the model’s ability to make accurate relative predictions. AR is computed by constructing a long-short portfolio, taking long positions in the top 10\% of stocks with the highest predicted returns and short positions in the bottom 10\%, thereby capturing returns from extreme predictions. The SR quantifies risk-adjusted performance by dividing the excess return (over a risk-free rate) by the standard deviation of returns. Together, these metrics provide a comprehensive view of both predictive quality and practical investment performance of our lead-lag detection framework.

\begin{table*}[!t]
\centering
\begin{minipage}{0.95\textwidth}
\centering
\caption{Comprehensive comparison of DeltaLag with baseline methods across different datasets.
Lead-lag indicates whether the model captures cross-asset or self-lead-lag relationships.
Delta refers to whether the lag value is dynamically identifiable on a daily basis. Specifically, learnable means the lag value is learned by the neural network during training.
Fixed indicates a constant lag value of 1. Offline means lag values are precomputed using statistical methods prior to training. Aggregated denotes that lead-lag effects are aggregated over multiple lag values. Adaptive indicates whether the model can dynamically learn which leader stocks provide predictive signals for the target stock.}
\resizebox{0.98\linewidth}{!}{
\begin{tabular}{r|ccc|ccc|ccc|ccc}
\toprule
\multirow{2}{*}{\centering\textbf{Method}} & \multirow{2}{*}{\centering\textbf{Lead-lag}} & \multirow{2}{*}{\centering\textbf{Delta}} & \multirow{2}{*}{\centering\textbf{Adaptive}} & \multicolumn{3}{c|}{\textbf{SP500}} & \multicolumn{3}{c|}{\textbf{NASDAQ}} & \multicolumn{3}{c}{\textbf{NYSE}} \\
& & & & \textbf{IC} & \textbf{AR} & \textbf{SR} & \textbf{IC} & \textbf{AR} & \textbf{SR} & \textbf{IC} & \textbf{AR} & \textbf{SR} \\
\midrule
DeltaLag (Ours) & Cross & \checkmark (learnable) & \checkmark & \textcolor{firstplace}{\textbf{0.0261}} & \textcolor{firstplace}{\textbf{0.2472}} & \textcolor{firstplace}{\textbf{2.1177}} & \textcolor{firstplace}{\textbf{0.0279}} & \textcolor{firstplace}{\textbf{0.3333}} & \textcolor{firstplace}{\textbf{2.9083}} & \textcolor{firstplace}{\textbf{0.0197}} & \textcolor{firstplace}{\textbf{0.2301}} & \textcolor{firstplace}{\textbf{2.5662}} \\
Lag1Net & Cross & \texttimes (fixed) & \checkmark & \textcolor{thirdplace}{0.0221} & 0.1662 & 1.2963 & \textcolor{thirdplace}{0.02206} & \textcolor{thirdplace}{0.2587} & 0.2695 & 0.0167 & 0.1458 & 1.3960 \\
SelfLagNet & Self & \checkmark (learnable) & \checkmark & \textcolor{thirdplace}{0.0221} & \textcolor{secondplace}{0.1887} & \textcolor{secondplace}{1.5640} & 0.0191 & 0.2552 & \textcolor{thirdplace}{2.2592} & \textcolor{thirdplace}{0.0174} & \textcolor{thirdplace}{0.1739} & \textcolor{thirdplace}{1.8109} \\
SelfLag1 & Self & \texttimes (fixed) & \texttimes & \textcolor{secondplace}{0.0233} & \textcolor{thirdplace}{0.1721} & \textcolor{thirdplace}{1.4533} & \textcolor{secondplace}{0.0250} & 0.2475 & \textcolor{thirdplace}{2.3070} & \textcolor{secondplace}{0.0194} & \textcolor{secondplace}{0.2014} & \textcolor{secondplace}{2.0735} \\
LagAll CorrGraph & Cross & \checkmark (offline) & \texttimes & 0.0054 & 0.1192 & 0.7703 & 0.0196 & \textcolor{secondplace}{0.2912} & \textcolor{secondplace}{2.3915} & -0.0001 & 0.0258 & -0.2658 \\
Lag1 CorrGraph & Cross & \texttimes (fixed) & \texttimes & 0.0059 & 0.0917 & 0.4222 & 0.0140 & 0.2548 & 1.8441 & 0.0001 & -0.0259 & -0.8772 \\
Meta-flow Graph & Cross & \texttimes (Aggregated) & \texttimes & -- & 0.0552 & 0.2804 & -- & 0.1248 & 0.8127 & -- & 0.0653 & 0.5808 \\
\bottomrule
\end{tabular}
}
\label{tab:baseline_comparison}
\end{minipage}
\end{table*}

\subsection{Baselines}
We categorize our baseline methods into multiple dimensions to comprehensively evaluate the effectiveness of DeltaLag in modeling lead-lag relationships. Specifically, we focus on the following four research questions:

\textit{(Q1) Does DeltaLag perform better by dynamically learning varying lag values across stock pairs than by using a fixed lag?}
We evaluate this by comparing DeltaLag against Lag1Net, a simplified variant of DeltaLag that removes the temporal dimension from the attention mechanism. Instead of attending over the past $l_{\text{max}}$ lags, Lag1Net uses only the final timestep (i.e., lag=1) for each peer stock, resulting in attention score matrices of dimension $\mathbb{R}^{(|\mathcal{S}| - 1)}$ instead of $\mathbb{R}^{(|\mathcal{S}| - 1) \times l_{\text{max}}}$. Lag1Net does not allow for dynamic lag discovery, serving as a fixed-lag baseline.

\textit{(Q2) Does modeling cross-asset lead-lag outperform self-lead-lag modeling?}
We compare DeltaLag against two self-only baselines: SelfLagNet and SelfLag1. 
Notably, DeltaLag excludes the target stock’s own sequence and relies solely on other stocks as potential leaders.
In essence, modeling cross- versus self-lead-lag relationships has distinct economic meanings. A model incorporating both might identify some stocks' leaders as themselves, thereby capturing their own internal dynamics like momentum or reversal.
In contrast, DeltaLag's cross-only approach offers a purer and more direct measure of the lead-lag effect, as its predictive power stems solely from external, inter-stock influences. To this end, we compare DeltaLag against two self-lead-lag models.
SelfLagNet adopts the same attention-based structure as DeltaLag but computes attention exclusively over the past $l_{\text{max}}$ timesteps of the target stock’s own history. SelfLag1 further simplifies this setup by using only the target stock’s features from the previous day (i.e., lag = 1), without any temporal attention. This comparison is designed to assess whether predictive signals from cross-stock leaders are more informative than those derived from the target stock itself.

\textit{(Q3) Can adaptively learned lead-lag relationships outperform those derived from statistical methods?}
We compare DeltaLag, which features an adaptive learning mechanism for lead-lag detection, with three baselines (LagAll CorrGraph, Lag1 CorrGraph, and Meta-flow Graph) that construct precomputed lead-lag graphs based on statistical methods. All methods model cross-asset lead-lag dependencies, but differ in how they obtain lead-lag pairs and assign lag values. LagAll CorrGraph constructs 10 Pearson correlation graphs over lags from 1 to 10 using historical data and selects top-$k$ leaders for each stock in a dynamic but offline manner. Lag1 CorrGraph simplifies this by precomputing correlations only at lag = 1. Meta-flow Graph is from Bennett et al.~\cite{BennettStefanos}, which represents the state-of-the-art approach for lead-lag detection. It quantifies lagged dependencies using distance correlation and CCF-AUC, then aggregates lead-lag effects over multiple lags and applies Hermitian random walk clustering to form a meta-flow graph.
In contrast, DeltaLag adaptively learns both stock leaders and optimal lag values on a daily basis without relying on any precomputed graph. This comparison allows us to evaluate whether DeltaLag's adaptive learning mechanism can outperform methods that rely on statistical-based lead-lag structures.

\textit{(Q4) Can our more interpretable lead-lag model achieve better performance than sophisticated deep learning models?}
We evaluate our framework against a range of strong baselines, including both temporal and spatio-temporal deep learning models for stock prediction or time series forecasting. This comparison allows us to assess whether our lead-lag detection model, while offering stronger interpretability than sophisticated "black box" deep learning models, can simultaneously achieve superior economic benefits.

\begin{figure*}[!t]
\centering
\caption{Cumulative returns comparison of our lead-lag model against various temporal and spatio-temporal models.}
\includegraphics[width=0.92\linewidth]{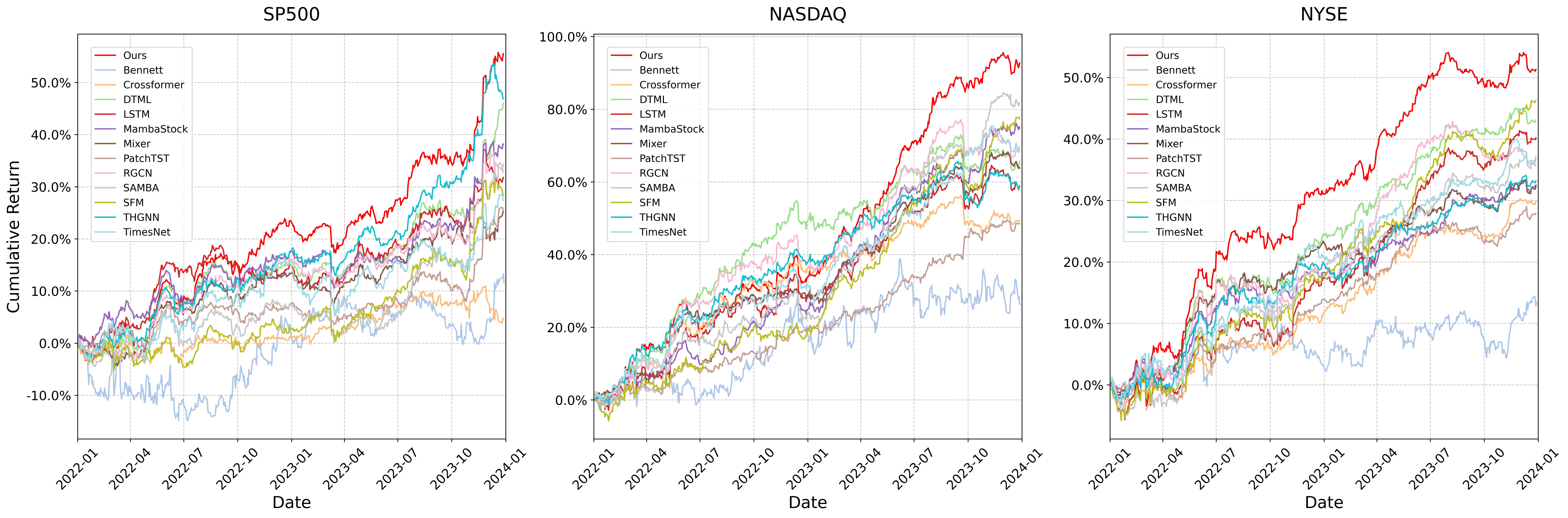}
\label{fig:cumulative_returns}
\end{figure*}

\begin{table*}[!t]
\centering
\begin{minipage}{0.87\textwidth}
\centering
\caption{Comparison of price-volume features vs. temporal embeddings as prediction inputs across different sequence models.}
\resizebox{0.97\linewidth}{!}{
\begin{tabular}{r|ccc|ccc|ccc}
\toprule
\multirow{2}{*}{\centering\textbf{Method}} & \multicolumn{3}{c|}{\textbf{SP500}} & \multicolumn{3}{c|}{\textbf{NASDAQ}} & \multicolumn{3}{c}{\textbf{NYSE}} \\
& \textbf{IC} & \textbf{AR} & \textbf{SR} & \textbf{IC} & \textbf{AR} & \textbf{SR} & \textbf{IC} & \textbf{AR} & \textbf{SR} \\
\midrule
LSTM + Raw Features (Ours) & \textbf{0.0261} & \textbf{0.2472} & \textbf{2.1177} & \textbf{0.0279} & \textbf{0.3333} & \textbf{2.9083} & 0.0197 & \textbf{0.2301} & \textbf{2.5662} \\
LSTM + Temporal Embeddings & 0.0227 & 0.1889 & 1.4829 & 0.0249 & 0.3211 & 2.4599 & \textbf{0.0208} & 0.2298 & 2.2069 \\
Mamba + Raw Features & 0.0260 & \textbf{0.1950} & \textbf{1.8197} & \textbf{0.0381} & \textbf{0.3672} & \textbf{3.4874} & \textbf{0.0261} & \textbf{0.1795} & \textbf{2.1901} \\
Mamba + Temporal Embeddings & \textbf{0.0327} & 0.1938 & 1.7071 & 0.0296 & 0.3577 & 3.1409 & 0.0178 & 0.1588 & 1.6915 \\
\bottomrule
\end{tabular}
}
\label{table:feature_comparison}
\end{minipage}
\end{table*}

\section{Results}
\subsection{Comparison with baseline models}
Table~\ref{tab:baseline_comparison} presents our experimental results, organized around the first three research questions outlined in our baseline evaluation.

To answer Question 1, we compare DeltaLag with Lag1Net, a simplified variant that removes the temporal dimension and assumes a fixed lag of 1 for all stock pairs. This setup allows us to isolate the contribution of our learnable delta design. The experimental results demonstrate that DeltaLag consistently outperforms Lag1Net across all metrics, indicating that different stock pairs require different optimal lag values. This finding validates that the ability to dynamically and adaptively learn varying lag values across different stock pairs provides significant predictive advantages over fixed lag assumptions.

For Question 2, our results strongly validate the superiority of our cross-asset lead-lag modeling over self-lead-lag approaches. Without relying on the target stock's own historical data in the prediction stage, DeltaLag outperforms both SelfLagNet and SelfLag1 across all evaluation metrics. This finding confirms that predictive signals from cross-stock leaders are more informative than those derived from the target stock's temporal patterns, demonstrating that inter-stock relationships are primarily crucial for effective modeling in financial markets.

Regarding Question 3, we examine whether adaptive learning of lead-lag relationships is more effective than relying on precomputed lead-lag graphs based on statistical methods. We compare DeltaLag against LagAll CorrGraph, Lag1 CorrGraph, and Meta-flow Graph. DeltaLag substantially outperforms all of them across all metrics. This significant performance gap between our adaptive model and the precomputed graph baselines indicates that lead-lag relationships observed in historical data do not reliably persist into the future. Lead-lag relationships identified through statistical methods fail to capture the dynamic, non-stationary nature of financial markets, where inter-stock dependencies continuously evolve over time. Our adaptive approach, which dynamically identifies lead-lag relationships at each timestep, effectively addresses this challenge by adapting to changing market conditions rather than relying on historical correlation patterns. 
When considering each potential leader's historical data (represented as a "key" with different lags), the primary advantage lies in our model's ability not only to select relevant target stocks (queries) based on similarity, but also to learn whether a given "key" at a specific lag is a strong predictor of the query’s next-day return. This process essentially allows the model to evaluate if an observed lead-lag correlation still possesses predictive momentum under current market conditions.

For Question 4, we conduct a comprehensive evaluation of our framework against a suite of temporal and spatio-temporal models for stock prediction and time series forecasting. These models include LSTM \cite{hochreiter_long_1997}, Crossformer \cite{zhang_crossformer_2023}, SFM \cite{zhang_stock_2017}, TimesNet \cite{wu_timesnet_2023}, MLP-Mixer \cite{tolstikhin_mlp-mixer_2021}, RGCN \cite{schlichtkrull_modeling_2018}, Bennett et al.'s model \cite{BennettStefanos}, THGNN \cite{xiang_temporal_2022}, PatchTST \cite{nie_time_2022}, MambaStock \cite{shi_mambastock_2024}, DTML \cite{yoo_accurate_2021}, and SAMBA \cite{ren_samba_2024}. 
To ensure a fair and rigorous comparison, all baseline models were also trained using our proposed ranking-based loss function instead of the MSE loss. \footnote{We observed in our experiments that the baseline models also achieved significantly better performance with the ranking-based loss compared to a standard MSE loss.}
The goal of this comparison is to assess whether our lead-lag detection model, while offering stronger interpretability, can simultaneously achieve superior economic benefits compared to sophisticated ``black box" deep learning models. Figure \ref{fig:cumulative_returns} illustrates the cumulative returns of our model alongside various established benchmarks over the test period.
As shown in Figure \ref{fig:cumulative_returns}, DeltaLag (marked as ``Ours'') consistently demonstrates a strong and stable upward trend in cumulative returns throughout the testing period. Quantitatively, our method attains an average of approximately 10 basis points (bpts) per day across the three datasets, surpassing all benchmark models in sustained profitability. This performance is practically significant, as it substantially exceeds typical transaction costs of 2-5 bpts for fees and market impact. This superior performance highlights a key insight: the lead-lag structures identified by our model encapsulate highly informative cross-stock dependencies that enable effective prediction even with straightforward downstream models. The ability to extract such economically valuable signals directly from raw price-volume data underscores the strength of our framework, offering both strong predictive performance and clear interpretability.

\begin{table*}[!t]
\centering
\begin{minipage}{0.85\textwidth}
\centering
\caption{Feature robustness analysis: Performance comparison using return-only features.}
\resizebox{0.99\linewidth}{!}{
\begin{tabular}{r|ccc|ccc|ccc}
\toprule
\multirow{2}{*}{\centering\textbf{Method}} & \multicolumn{3}{c|}{\textbf{SP500}} & \multicolumn{3}{c|}{\textbf{NASDAQ}} & \multicolumn{3}{c}{\textbf{NYSE}} \\
& \textbf{IC} & \textbf{AR} & \textbf{SR} & \textbf{IC} & \textbf{AR} & \textbf{SR} & \textbf{IC} & \textbf{AR} & \textbf{SR} \\
\midrule
DeltaLag (Ours) & \textbf{0.0211} & \textbf{0.1576} & \textbf{1.1529} & \textbf{0.0251} & \textbf{0.3918} & \textbf{3.2919} & \textbf{0.0208} & \textbf{0.2354} & \textbf{2.434} \\
Lag1Net & 0.0083 & 0.048 & 0.2086 & 0.0036 & 0.0015 & -0.4574 & 0.0132 & 0.037 & -0.0515 \\
SelfLagNet & 0.0133 & 0.0967 & 0.5447 & 0.0181 & 0.2695 & 1.7657 & 0.0151 & 0.1847 & 1.5229 \\
SelfLag1 & 0.0059 & 0.0343 & -0.0462 & -0.0032 & -0.18 & -1.517 & 0.0028 & -0.0302 & -0.6277 \\
\bottomrule
\end{tabular}
}
\label{table:robustness_analysis}
\end{minipage}
\end{table*}

\begin{table*}[!t]
\centering
\begin{minipage}{0.83\textwidth}
\centering
\caption{Comparison of different loss functions across datasets.}
\resizebox{0.99\linewidth}{!}{
\begin{tabular}{r|ccc|ccc|ccc}
\toprule
\multirow{2}{*}{\centering\textbf{Loss function}} & \multicolumn{3}{c|}{\textbf{SP500}} & \multicolumn{3}{c|}{\textbf{NASDAQ}} & \multicolumn{3}{c}{\textbf{NYSE}} \\
& \textbf{IC} & \textbf{AR} & \textbf{SR} & \textbf{IC} & \textbf{AR} & \textbf{SR} & \textbf{IC} & \textbf{AR} & \textbf{SR} \\
\midrule
Ours & 0.0261 & \textbf{0.2472} & \textbf{2.1177} & \textbf{0.0279} & \textbf{0.3333} & \textbf{2.9083} & \textbf{0.0197} & \textbf{0.2301} & \textbf{2.5662} \\
ICLoss & \textbf{0.0459} & 0.2205 & 1.9402 & 0.0211 & 0.2496 & 1.9709 & 0.0192 & 0.1192 & 0.2779 \\
MSELoss & -0.0008 & -0.0068 & -0.7513 & 0.0019 & 0.0272 & -0.227 & -0.0015 & -0.0183 & -1.5344 \\
\bottomrule
\end{tabular}
}
\label{table:loss_comparison}
\end{minipage}
\end{table*}

\subsection{Ablation studies}
To further investigate the effectiveness of our model design choices, we conduct comprehensive ablation studies examining three key aspects of our framework.

The first aspect examines the choice of input features for the prediction stage. After identifying lead-lag relationships through our cross-attention mechanism, we have two options for the inputs to our prediction network (a simple MLP):

1. Raw Feature Prediction: Use the original price-volume features of the leading stocks at their identified lag times:
\begin{equation}
    \mathbf{z}_{v_i,u,t} = \mathbf{x}_{v_i,t-\tau_{v_i,u}} \in \mathbb{R}^F
\end{equation}

2. Temporal Embedding Prediction: Use the hidden states of the leading stocks obtained from the sequence model at their identified lag times: 
\begin{equation}
    \mathbf{z}_{v_i,u,t} = \mathbf{X}'_{v_i,t}[L-\tau_{v_i,u},:] \in \mathbb{R}^N
\end{equation}
Table \ref{table:feature_comparison} presents the results of this comparison across two different sequence models on the test set. The results clearly demonstrate that once lead-lag pairs and their corresponding lag values are identified, using raw price-volume features of the leading stocks yields superior economic performance compared to using temporal embeddings, regardless of the sequence model employed. This finding indicates that the original daily features provide stronger predictive signals than the aggregated temporal representations learned by sequence models. By utilizing raw features, our model not only achieves better economic results but also significantly enhances interpretability, as it directly identifies specific trading days and stocks whose original features contribute to the prediction, rather than relying on complex, potentially opaque temporal embeddings.

The second aspect evaluates the robustness of our lead-lag detection framework by conducting experiments using limited feature sets to assess whether our model can maintain effectiveness even with minimal information. Specifically, we compare the performance when using only single-day returns, calculated solely from stock closing prices, as the only input feature. This analysis is crucial for understanding the robustness of our approach in scenarios where feature availability is constrained. Table \ref{table:robustness_analysis} shows the results of this robustness analysis. The performance advantage becomes more pronounced when using only single-day return data. In this more challenging scenario with limited information, our return-only adaptive lead-lag model maintains strong performance while some other baseline methods experience dramatic performance degradation. 
This robustness analysis confirms that the effectiveness of our framework does not rely on rich feature sets, but rather stems from its ability to accurately identify lead-lag relationships and generate economic benefits even with limited features.

The third aspect investigates the impact of different loss functions on model performance. Table \ref{table:loss_comparison} presents a comparison of our proposed loss function against two alternatives: Information Coefficient (IC) Loss and Mean Squared Error (MSE) Loss. The results demonstrate that our proposed loss function achieves superior economic benefits with higher AR and SR. Meanwhile, MSE Loss performs poorly across all metrics, highlighting the importance of using ranking-based objectives for financial prediction tasks. This confirms that our loss function design effectively optimizes for the economic metrics that matter most in practical trading applications.

\subsection{Lead-lag property}
\begin{figure}[tbp]
    \centering
    \includegraphics[width=0.40\textwidth]{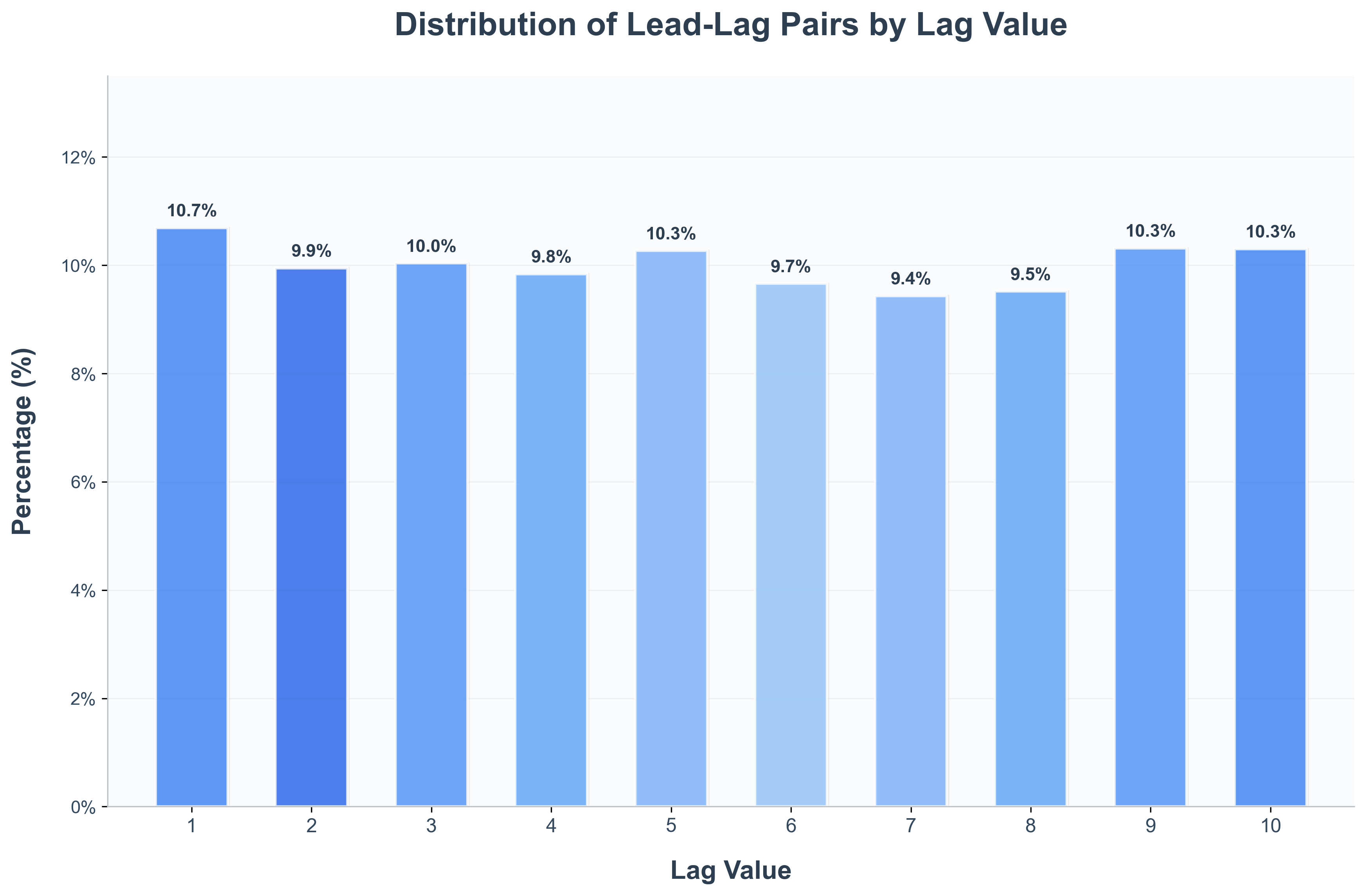}
    \caption{Distribution of Lead-Lag pairs by lag value}
    \label{fig:lag_distribution}
\end{figure}

To gain deeper insights into the lead-lag relationships identified by our model, we analyze the patterns discovered during the test period.  We first analyze the distribution of the lag values. Figure \ref{fig:lag_distribution} shows that the distribution of lag values is remarkably uniform, with each lag value from 1 to 9 days accounting for approximately 10-12\% of all identified lead-lag relationships. This suggests that meaningful lead-lag relationships exist across various time horizons rather than being concentrated at specific lags.

Although the approximately uniform pattern may appear counterintuitive at first glance and seem to contradict standard financial intuition,  it does not necessarily imply that the model assigns lag values randomly or  merely reflects noise. Instead, we interpret it as evidence that the model dynamically adapts to changing market conditions. At any given timestep, the attention mechanism may emphasize short-, medium-, or long-horizon lags depending on prevailing market conditions, and when aggregating over the full period, these dynamically-changing preferences yield an overall flat distribution. In this interpretation, the resulting uniformity indicates flexibility and diversification of lag relationships rather than an absence of structure. 


Our analysis also reveals interesting clustering patterns in the identified lead-lag relationships. For example, across the S\&P 500 stocks, we observe 38.18 unique leader stocks per date out of 500 total stocks, indicating significant concentration in leadership roles. When considering only the highest-ranked lead-lag pairs (rank 1 by attention score), this concentration becomes even more pronounced, with an average of just 22.14 unique leader stocks per date. This emergent clustering property aligns well with traditional statistical-based lead-lag model like Bennett et al.'s approach \cite{BennettStefanos}, which explicitly uses Hermitian RW clustering to form a meta-flow graph among clusters. Remarkably, our neural network model automatically identifies these stock clusters through the attention mechanism without being explicitly designed to do so.

\section{Conclusion}
In this paper, we introduced a novel deep learning method for lead-lag relationship detection in financial markets. Our model employs a cross-attention mechanism and a tailored ranking loss function to adaptively identify lead-lag pairs and their corresponding lag values, followed by a simple MLP for return prediction. Through extensive empirical evaluation, we demonstrated that our approach consistently outperforms baselines that rely on a fixed lag value or self-lead-lag modeling. Our findings also confirm the weak momentum property of correlation-based lead-lag relationships derived from statistical methods, and highlight the importance of constructing adaptive graphs that can dynamically adjust to evolving market conditions. Finally, DeltaLag surpasses a wide range of temporal and spatio-temporal deep learning models designed for stock prediction or time series forecasting tasks, achieving substantially higher trading profitability while maintaining stronger economic rationale and interpretability.

While performing strongly, the reliance of our model on daily price-volume data suggests further research extensions. Future work could include integrating diverse data sources such as macroeconomic indicators, exploring an intraday version of this model to capture higher-frequency dynamics, and incorporating factor models by jointly inferring both the exposure matrix and the corresponding lead-lag matrix for each stock against various factors. 

 \begin{acks}
This work is supported by the Guangzhou-HKUST(GZ) Joint Funding Program (No. 2024A03J0630).
\end{acks}

\bibliographystyle{ACM-Reference-Format}
\bibliography{sample-base}


\begin{thebibliography}{34}


\ifx \showCODEN    \undefined \def \showCODEN     #1{\unskip}     \fi
\ifx \showISBNx    \undefined \def \showISBNx     #1{\unskip}     \fi
\ifx \showISBNxiii \undefined \def \showISBNxiii  #1{\unskip}     \fi
\ifx \showISSN     \undefined \def \showISSN      #1{\unskip}     \fi
\ifx \showLCCN     \undefined \def \showLCCN      #1{\unskip}     \fi
\ifx \shownote     \undefined \def \shownote      #1{#1}          \fi
\ifx \showarticletitle \undefined \def \showarticletitle #1{#1}   \fi
\ifx \showURL      \undefined \def \showURL       {\relax}        \fi
\providecommand\bibfield[2]{#2}
\providecommand\bibinfo[2]{#2}
\providecommand\natexlab[1]{#1}
\providecommand\showeprint[2][]{arXiv:#2}

\bibitem[Albers et~al\mbox{.}(2021)]%
        {albers2021fragmentation}
\bibfield{author}{\bibinfo{person}{Jakob Albers}, \bibinfo{person}{Mihai Cucuringu}, \bibinfo{person}{Sam Howison}, {and} \bibinfo{person}{Alexander~Y Shestopaloff}.} \bibinfo{year}{2021}\natexlab{}.
\newblock \showarticletitle{Fragmentation, price formation and cross-impact in bitcoin markets}.
\newblock \bibinfo{journal}{\emph{Applied Mathematical Finance}} \bibinfo{volume}{28}, \bibinfo{number}{5} (\bibinfo{year}{2021}), \bibinfo{pages}{395--448}.
\newblock


\bibitem[Bennett et~al\mbox{.}(2022)]%
        {BennettStefanos}
\bibfield{author}{\bibinfo{person}{Stefanos Bennett}, \bibinfo{person}{Mihai Cucuringu}, {and} \bibinfo{person}{Gesine Reinert}.} \bibinfo{year}{2022}\natexlab{}.
\newblock \showarticletitle{Lead--lag detection and network clustering for multivariate time series with an application to the US equity market}.
\newblock \bibinfo{journal}{\emph{Machine Learning}} \bibinfo{volume}{111}, \bibinfo{number}{12} (\bibinfo{year}{2022}), \bibinfo{pages}{4497--4538}.
\newblock


\bibitem[Buccheri et~al\mbox{.}(2021)]%
        {buccheri2021high}
\bibfield{author}{\bibinfo{person}{Giuseppe Buccheri}, \bibinfo{person}{Fulvio Corsi}, {and} \bibinfo{person}{Stefano Peluso}.} \bibinfo{year}{2021}\natexlab{}.
\newblock \showarticletitle{High-frequency lead-lag effects and cross-asset linkages: a multi-asset lagged adjustment model}.
\newblock \bibinfo{journal}{\emph{Journal of Business \& Economic Statistics}} \bibinfo{volume}{39}, \bibinfo{number}{3} (\bibinfo{year}{2021}), \bibinfo{pages}{605--621}.
\newblock


\bibitem[Cartea et~al\mbox{.}(2023)]%
        {qiLeadLag}
\bibfield{author}{\bibinfo{person}{Álvaro Cartea}, \bibinfo{person}{Mihai Cucuringu}, {and} \bibinfo{person}{Qi Jin}.} \bibinfo{year}{2023}\natexlab{}.
\newblock \showarticletitle{Detecting Lead-Lag Relationships in Stock Returns and Portfolio Strategies}.
\newblock  (\bibinfo{date}{October} \bibinfo{year}{2023}).
\newblock
\newblock
\shownote{Available at SSRN: \url{https://ssrn.com/abstract=4599565}}.


\bibitem[Cucuringu et~al\mbox{.}(2020)]%
        {HermitianClust}
\bibfield{author}{\bibinfo{person}{M. Cucuringu}, \bibinfo{person}{H. Li}, \bibinfo{person}{H. Sun}, {and} \bibinfo{person}{L. Zanetti}.} \bibinfo{year}{2020}\natexlab{}.
\newblock \showarticletitle{Hermitian matrices for clustering directed graphs: insights and applications}. In \bibinfo{booktitle}{\emph{Proceedings of the 23rd International Conference on Artificial Intelligence and Statistics (AISTATS 2020)}}.
\newblock


\bibitem[De~Luca and Pizzolante(2021)]%
        {DeLucaGiovanni}
\bibfield{author}{\bibinfo{person}{Giovanni De~Luca} {and} \bibinfo{person}{Federica Pizzolante}.} \bibinfo{year}{2021}\natexlab{}.
\newblock \showarticletitle{Detecting leaders country from road transport emission time-series}.
\newblock \bibinfo{journal}{\emph{Environments}} \bibinfo{volume}{8}, \bibinfo{number}{3} (\bibinfo{year}{2021}), \bibinfo{pages}{18}.
\newblock


\bibitem[Fan et~al\mbox{.}(2022)]%
        {fan_does_2022}
\bibfield{author}{\bibinfo{person}{Ningyuan Fan}, \bibinfo{person}{Zhi-Ping Fan}, \bibinfo{person}{Yongli Li}, {and} \bibinfo{person}{Meng Li}.} \bibinfo{year}{2022}\natexlab{}.
\newblock \showarticletitle{Does the lead-lag effect exist in stock markets?}
\newblock \bibinfo{journal}{\emph{Applied Economics Letters}} \bibinfo{volume}{29}, \bibinfo{number}{10} (\bibinfo{date}{June} \bibinfo{year}{2022}), \bibinfo{pages}{895--900}.
\newblock
\showISSN{1350-4851}
\href{https://doi.org/10.1080/13504851.2021.1897068}{doi:\nolinkurl{10.1080/13504851.2021.1897068}}
\newblock
\shownote{Publisher: Routledge \_eprint: https://doi.org/10.1080/13504851.2021.1897068}.


\bibitem[Feng et~al\mbox{.}(2019)]%
        {feng2019temporal}
\bibfield{author}{\bibinfo{person}{Fuli Feng}, \bibinfo{person}{Xiangnan He}, \bibinfo{person}{Xiang Wang}, \bibinfo{person}{Cheng Luo}, \bibinfo{person}{Yiqun Liu}, {and} \bibinfo{person}{Tat-Seng Chua}.} \bibinfo{year}{2019}\natexlab{}.
\newblock \showarticletitle{Temporal relational ranking for stock prediction}.
\newblock \bibinfo{journal}{\emph{ACM Transactions on Information Systems (TOIS)}} \bibinfo{volume}{37}, \bibinfo{number}{2} (\bibinfo{year}{2019}), \bibinfo{pages}{1--30}.
\newblock


\bibitem[Hochreiter and Schmidhuber(1997)]%
        {hochreiter_long_1997}
\bibfield{author}{\bibinfo{person}{Sepp Hochreiter} {and} \bibinfo{person}{Jürgen Schmidhuber}.} \bibinfo{year}{1997}\natexlab{}.
\newblock \showarticletitle{Long short-term memory}.
\newblock \bibinfo{journal}{\emph{Neural computation}} \bibinfo{volume}{9}, \bibinfo{number}{8} (\bibinfo{year}{1997}), \bibinfo{pages}{1735--1780}.
\newblock
\newblock
\shownote{Publisher: MIT Press}.


\bibitem[Li et~al\mbox{.}(2021)]%
        {LiYongli}
\bibfield{author}{\bibinfo{person}{Yongli Li}, \bibinfo{person}{Chao Liu}, \bibinfo{person}{Tianchen Wang}, {and} \bibinfo{person}{Baiqing Sun}.} \bibinfo{year}{2021}\natexlab{}.
\newblock \showarticletitle{Dynamic patterns of daily lead-lag networks in stock markets}.
\newblock \bibinfo{journal}{\emph{Quantitative Finance}} \bibinfo{volume}{21}, \bibinfo{number}{12} (\bibinfo{year}{2021}), \bibinfo{pages}{2055--2068}.
\newblock


\bibitem[Li et~al\mbox{.}(2017)]%
        {li2017improving}
\bibfield{author}{\bibinfo{person}{Yuncheng Li}, \bibinfo{person}{Yale Song}, {and} \bibinfo{person}{Jiebo Luo}.} \bibinfo{year}{2017}\natexlab{}.
\newblock \showarticletitle{Improving pairwise ranking for multi-label image classification}. In \bibinfo{booktitle}{\emph{Proceedings of the IEEE conference on computer vision and pattern recognition}}. \bibinfo{pages}{3617--3625}.
\newblock


\bibitem[Li et~al\mbox{.}(2024)]%
        {li2024stock}
\bibfield{author}{\bibinfo{person}{Zhichao Li}, \bibinfo{person}{Xianghui Yuan}, \bibinfo{person}{Liwei Jin}, {and} \bibinfo{person}{Chencheng Zhao}.} \bibinfo{year}{2024}\natexlab{}.
\newblock \showarticletitle{Stock trend prediction: an effective hybrid deep model based on lead and lag correlation graphs}. In \bibinfo{booktitle}{\emph{2024 27th International Conference on Computer Supported Cooperative Work in Design (CSCWD)}}. IEEE, \bibinfo{pages}{103--108}.
\newblock


\bibitem[Miori and Cucuringu(2022)]%
        {MioriDeborah}
\bibfield{author}{\bibinfo{person}{Deborah Miori} {and} \bibinfo{person}{Mihai Cucuringu}.} \bibinfo{year}{2022}\natexlab{}.
\newblock \showarticletitle{Returns-Driven Macro Regimes and Characteristic Lead-Lag Behaviour between Asset Classes}.
\newblock \bibinfo{journal}{\emph{arXiv preprint arXiv:2209.00268}} (\bibinfo{year}{2022}).
\newblock


\bibitem[Nauta et~al\mbox{.}(2019)]%
        {nauta2019causal}
\bibfield{author}{\bibinfo{person}{Meike Nauta}, \bibinfo{person}{Doina Bucur}, {and} \bibinfo{person}{Christin Seifert}.} \bibinfo{year}{2019}\natexlab{}.
\newblock \showarticletitle{Causal discovery with attention-based convolutional neural networks}.
\newblock \bibinfo{journal}{\emph{Machine Learning and Knowledge Extraction}} \bibinfo{volume}{1}, \bibinfo{number}{1} (\bibinfo{year}{2019}), \bibinfo{pages}{19}.
\newblock


\bibitem[Nie et~al\mbox{.}(2022)]%
        {nie_time_2022}
\bibfield{author}{\bibinfo{person}{Yuqi Nie}, \bibinfo{person}{Nam~H. Nguyen}, \bibinfo{person}{Phanwadee Sinthong}, {and} \bibinfo{person}{Jayant Kalagnanam}.} \bibinfo{year}{2022}\natexlab{}.
\newblock \showarticletitle{A {Time} {Series} is {Worth} 64 {Words}: {Long}-term {Forecasting} with {Transformers}}.
\newblock  (\bibinfo{year}{2022}).
\newblock
\href{https://doi.org/10.48550/ARXIV.2211.14730}{doi:\nolinkurl{10.48550/ARXIV.2211.14730}}
\newblock
\shownote{Publisher: arXiv Version Number: 2}.


\bibitem[Ren et~al\mbox{.}(2024)]%
        {ren_samba_2024}
\bibfield{author}{\bibinfo{person}{Liliang Ren}, \bibinfo{person}{Yang Liu}, \bibinfo{person}{Yadong Lu}, \bibinfo{person}{Yelong Shen}, \bibinfo{person}{Chen Liang}, {and} \bibinfo{person}{Weizhu Chen}.} \bibinfo{year}{2024}\natexlab{}.
\newblock \showarticletitle{Samba: {Simple} {Hybrid} {State} {Space} {Models} for {Efficient} {Unlimited} {Context} {Language} {Modeling}}.
\newblock
\urldef\tempurl%
\url{https://openreview.net/forum?id=bIlnpVM4bc}
\showURL{%
\tempurl}


\bibitem[Runge et~al\mbox{.}(2019)]%
        {RungeJakob}
\bibfield{author}{\bibinfo{person}{Jakob Runge}, \bibinfo{person}{Peer Nowack}, \bibinfo{person}{Marlene Kretschmer}, \bibinfo{person}{Seth Flaxman}, {and} \bibinfo{person}{Dino Sejdinovic}.} \bibinfo{year}{2019}\natexlab{}.
\newblock \showarticletitle{Detecting and quantifying causal associations in large nonlinear time series datasets}.
\newblock \bibinfo{journal}{\emph{Science advances}} \bibinfo{volume}{5}, \bibinfo{number}{11} (\bibinfo{year}{2019}), \bibinfo{pages}{eaau4996}.
\newblock


\bibitem[Schlichtkrull et~al\mbox{.}(2018)]%
        {schlichtkrull_modeling_2018}
\bibfield{author}{\bibinfo{person}{Michael Schlichtkrull}, \bibinfo{person}{Thomas~N. Kipf}, \bibinfo{person}{Peter Bloem}, \bibinfo{person}{Rianne van~den Berg}, \bibinfo{person}{Ivan Titov}, {and} \bibinfo{person}{Max Welling}.} \bibinfo{year}{2018}\natexlab{}.
\newblock \showarticletitle{Modeling {Relational} {Data} with {Graph} {Convolutional} {Networks}}. In \bibinfo{booktitle}{\emph{The {Semantic} {Web}}} \emph{(\bibinfo{series}{Lecture {Notes} in {Computer} {Science}})}, \bibfield{editor}{\bibinfo{person}{Aldo Gangemi}, \bibinfo{person}{Roberto Navigli}, \bibinfo{person}{Maria-Esther Vidal}, \bibinfo{person}{Pascal Hitzler}, \bibinfo{person}{Raphaël Troncy}, \bibinfo{person}{Laura Hollink}, \bibinfo{person}{Anna Tordai}, {and} \bibinfo{person}{Mehwish Alam}} (Eds.). \bibinfo{publisher}{Springer International Publishing}, \bibinfo{address}{Cham}, \bibinfo{pages}{593--607}.
\newblock
\showISBNx{978-3-319-93417-4}
\href{https://doi.org/10.1007/978-3-319-93417-4_38}{doi:\nolinkurl{10.1007/978-3-319-93417-4_38}}


\bibitem[Shi et~al\mbox{.}(2023)]%
        {ShiDanni}
\bibfield{author}{\bibinfo{person}{Danni Shi}, \bibinfo{person}{Jan-Peter Calliess}, {and} \bibinfo{person}{Mihai Cucuringu}.} \bibinfo{year}{2023}\natexlab{}.
\newblock \showarticletitle{Multireference Alignment for Lead-Lag Detection in Multivariate Time Series and Equity Trading}. In \bibinfo{booktitle}{\emph{Proceedings of the Fourth ACM International Conference on AI in Finance}}. \bibinfo{pages}{507--515}.
\newblock


\bibitem[Shi(2024)]%
        {shi_mambastock_2024}
\bibfield{author}{\bibinfo{person}{Zhuangwei Shi}.} \bibinfo{year}{2024}\natexlab{}.
\newblock \bibinfo{title}{{MambaStock}: {Selective} state space model for stock prediction}.
\newblock
\href{https://doi.org/10.48550/arXiv.2402.18959}{doi:\nolinkurl{10.48550/arXiv.2402.18959}}
\newblock
\shownote{arXiv:2402.18959 [cs]}.


\bibitem[Tolikas(2018)]%
        {tolikas2018lead}
\bibfield{author}{\bibinfo{person}{Konstantinos Tolikas}.} \bibinfo{year}{2018}\natexlab{}.
\newblock \showarticletitle{The lead-lag relation between the stock and the bond markets}.
\newblock \bibinfo{journal}{\emph{The European Journal of Finance}} \bibinfo{volume}{24}, \bibinfo{number}{10} (\bibinfo{year}{2018}), \bibinfo{pages}{849--866}.
\newblock


\bibitem[Tolstikhin et~al\mbox{.}(2021)]%
        {tolstikhin_mlp-mixer_2021}
\bibfield{author}{\bibinfo{person}{Ilya Tolstikhin}, \bibinfo{person}{Neil Houlsby}, \bibinfo{person}{Alexander Kolesnikov}, \bibinfo{person}{Lucas Beyer}, \bibinfo{person}{Xiaohua Zhai}, \bibinfo{person}{Thomas Unterthiner}, \bibinfo{person}{Jessica Yung}, \bibinfo{person}{Andreas Steiner}, \bibinfo{person}{Daniel Keysers}, \bibinfo{person}{Jakob Uszkoreit}, \bibinfo{person}{Mario Lucic}, {and} \bibinfo{person}{Alexey Dosovitskiy}.} \bibinfo{year}{2021}\natexlab{}.
\newblock \showarticletitle{{MLP}-{Mixer}: {An} all-{MLP} {Architecture} for {Vision}}.
\newblock \bibinfo{journal}{\emph{arXiv:2105.01601 [cs]}} (\bibinfo{date}{June} \bibinfo{year}{2021}).
\newblock
\urldef\tempurl%
\url{http://arxiv.org/abs/2105.01601}
\showURL{%
\tempurl}
\newblock
\shownote{arXiv: 2105.01601}.


\bibitem[Wang et~al\mbox{.}(2025)]%
        {WangSaizhuo}
\bibfield{author}{\bibinfo{person}{Saizhuo Wang}, \bibinfo{person}{Hao Kong}, \bibinfo{person}{Jiadong Guo}, \bibinfo{person}{Fengrui Hua}, \bibinfo{person}{Yiyan Qi}, \bibinfo{person}{Wanyun Zhou}, \bibinfo{person}{Jiahao Zheng}, \bibinfo{person}{Xinyu Wang}, \bibinfo{person}{Lionel~M Ni}, {and} \bibinfo{person}{Jian Guo}.} \bibinfo{year}{2025}\natexlab{}.
\newblock \showarticletitle{QuantBench: Benchmarking AI Methods for Quantitative Investment}.
\newblock \bibinfo{journal}{\emph{arXiv preprint arXiv:2504.18600}} (\bibinfo{year}{2025}).
\newblock


\bibitem[Wu et~al\mbox{.}(2010)]%
        {WuDi}
\bibfield{author}{\bibinfo{person}{Di Wu}, \bibinfo{person}{Yiping Ke}, \bibinfo{person}{Jeffrey~Xu Yu}, \bibinfo{person}{Philip~S Yu}, {and} \bibinfo{person}{Lei Chen}.} \bibinfo{year}{2010}\natexlab{}.
\newblock \showarticletitle{Detecting leaders from correlated time series}. In \bibinfo{booktitle}{\emph{Database Systems for Advanced Applications: 15th International Conference, DASFAA 2010, Tsukuba, Japan, April 1-4, 2010, Proceedings, Part I 15}}. Springer, \bibinfo{pages}{352--367}.
\newblock


\bibitem[Wu et~al\mbox{.}(2023)]%
        {wu_timesnet_2023}
\bibfield{author}{\bibinfo{person}{Haixu Wu}, \bibinfo{person}{Tengge Hu}, \bibinfo{person}{Yong Liu}, \bibinfo{person}{Hang Zhou}, \bibinfo{person}{Jianmin Wang}, {and} \bibinfo{person}{Mingsheng Long}.} \bibinfo{year}{2023}\natexlab{}.
\newblock \bibinfo{title}{{TimesNet}: {Temporal} {2D}-{Variation} {Modeling} for {General} {Time} {Series} {Analysis}}.
\newblock
\href{https://doi.org/10.48550/arXiv.2210.02186}{doi:\nolinkurl{10.48550/arXiv.2210.02186}}
\newblock
\shownote{arXiv:2210.02186 [cs]}.


\bibitem[Xiang et~al\mbox{.}(2022)]%
        {xiang_temporal_2022}
\bibfield{author}{\bibinfo{person}{Sheng Xiang}, \bibinfo{person}{Dawei Cheng}, \bibinfo{person}{Chencheng Shang}, \bibinfo{person}{Ying Zhang}, {and} \bibinfo{person}{Yuqi Liang}.} \bibinfo{year}{2022}\natexlab{}.
\newblock \showarticletitle{Temporal and {Heterogeneous} {Graph} {Neural} {Network} for {Financial} {Time} {Series} {Prediction}}. In \bibinfo{booktitle}{\emph{Proceedings of the 31st {ACM} {International} {Conference} on {Information} \& {Knowledge} {Management}}} \emph{(\bibinfo{series}{{CIKM} '22})}. \bibinfo{publisher}{Association for Computing Machinery}, \bibinfo{address}{New York, NY, USA}, \bibinfo{pages}{3584--3593}.
\newblock
\showISBNx{978-1-4503-9236-5}
\href{https://doi.org/10.1145/3511808.3557089}{doi:\nolinkurl{10.1145/3511808.3557089}}


\bibitem[Yoo et~al\mbox{.}(2021)]%
        {yoo_accurate_2021}
\bibfield{author}{\bibinfo{person}{Jaemin Yoo}, \bibinfo{person}{Yejun Soun}, \bibinfo{person}{Yong-chan Park}, {and} \bibinfo{person}{U Kang}.} \bibinfo{year}{2021}\natexlab{}.
\newblock \showarticletitle{Accurate {Multivariate} {Stock} {Movement} {Prediction} via {Data}-{Axis} {Transformer} with {Multi}-{Level} {Contexts}}. In \bibinfo{booktitle}{\emph{Proceedings of the 27th {ACM} {SIGKDD} {Conference} on {Knowledge} {Discovery} \& {Data} {Mining}}} \emph{(\bibinfo{series}{{KDD} '21})}. \bibinfo{publisher}{Association for Computing Machinery}, \bibinfo{address}{New York, NY, USA}, \bibinfo{pages}{2037--2045}.
\newblock
\showISBNx{978-1-4503-8332-5}
\href{https://doi.org/10.1145/3447548.3467297}{doi:\nolinkurl{10.1145/3447548.3467297}}


\bibitem[Zhang et~al\mbox{.}(2017)]%
        {zhang_stock_2017}
\bibfield{author}{\bibinfo{person}{Liheng Zhang}, \bibinfo{person}{Charu Aggarwal}, {and} \bibinfo{person}{Guo-Jun Qi}.} \bibinfo{year}{2017}\natexlab{}.
\newblock \showarticletitle{Stock {Price} {Prediction} via {Discovering} {Multi}-{Frequency} {Trading} {Patterns}}. In \bibinfo{booktitle}{\emph{Proceedings of the 23rd {ACM} {SIGKDD} {International} {Conference} on {Knowledge} {Discovery} and {Data} {Mining}}}. \bibinfo{publisher}{ACM}, \bibinfo{address}{Halifax NS Canada}, \bibinfo{pages}{2141--2149}.
\newblock
\showISBNx{978-1-4503-4887-4}
\href{https://doi.org/10.1145/3097983.3098117}{doi:\nolinkurl{10.1145/3097983.3098117}}


\bibitem[Zhang et~al\mbox{.}(2023a)]%
        {ZhangYichiDynamic}
\bibfield{author}{\bibinfo{person}{Yichi Zhang}, \bibinfo{person}{Mihai Cucuringu}, \bibinfo{person}{Alexander Shestopaloff}, {and} \bibinfo{person}{Stefan Zohren}.} \bibinfo{year}{2023}\natexlab{a}.
\newblock \showarticletitle{Dynamic Time Warping for Lead-Lag Relationship Detection in Lagged Multi-Factor Models}. In \bibinfo{booktitle}{\emph{Proceedings of the Fourth ACM International Conference on AI in Finance}}. \bibinfo{pages}{454--462}.
\newblock


\bibitem[Zhang et~al\mbox{.}(2023b)]%
        {ZhangYichiRobust}
\bibfield{author}{\bibinfo{person}{Yichi Zhang}, \bibinfo{person}{Mihai Cucuringu}, \bibinfo{person}{Alexander~Y Shestopaloff}, {and} \bibinfo{person}{Stefan Zohren}.} \bibinfo{year}{2023}\natexlab{b}.
\newblock \showarticletitle{Robust Detection of Lead-Lag Relationships in Lagged Multi-Factor Models}.
\newblock \bibinfo{journal}{\emph{arXiv preprint arXiv:2305.06704}} (\bibinfo{year}{2023}).
\newblock


\bibitem[Zhang and Yan(2023)]%
        {zhang_crossformer_2023}
\bibfield{author}{\bibinfo{person}{Yunhao Zhang} {and} \bibinfo{person}{Junchi Yan}.} \bibinfo{year}{2023}\natexlab{}.
\newblock \showarticletitle{Crossformer: {Transformer} {Utilizing} {Cross}-{Dimension} {Dependency} for {Multivariate} {Time} {Series} {Forecasting}}.
\newblock
\urldef\tempurl%
\url{https://openreview.net/forum?id=vSVLM2j9eie}
\showURL{%
\tempurl}


\bibitem[Zheng et~al\mbox{.}(2023)]%
        {ZhengXiaolin}
\bibfield{author}{\bibinfo{person}{Xiaolin Zheng}, \bibinfo{person}{Mengpu Liu}, {and} \bibinfo{person}{Mengying Zhu}.} \bibinfo{year}{2023}\natexlab{}.
\newblock \showarticletitle{Deep Hashing-based Dynamic Stock Correlation Estimation via Normalizing Flow.}. In \bibinfo{booktitle}{\emph{IJCAI}}. \bibinfo{pages}{4993--5001}.
\newblock


\bibitem[Zhong et~al\mbox{.}(2024)]%
        {zhong_dspo_2024}
\bibfield{author}{\bibinfo{person}{Jianyuan Zhong}, \bibinfo{person}{Zhijian Xu}, \bibinfo{person}{Saizhuo Wang}, \bibinfo{person}{Xiangyu Wen}, \bibinfo{person}{Jian Guo}, {and} \bibinfo{person}{Qiang Xu}.} \bibinfo{year}{2024}\natexlab{}.
\newblock \bibinfo{title}{{DSPO}: {An} {End}-to-{End} {Framework} for {Direct} {Sorted} {Portfolio} {Construction}}.
\newblock
\href{https://doi.org/10.48550/arXiv.2405.15833}{doi:\nolinkurl{10.48550/arXiv.2405.15833}}
\newblock
\shownote{arXiv:2405.15833 [q-fin]}.


\bibitem[Zhu et~al\mbox{.}(2022)]%
        {ZhuHaoren}
\bibfield{author}{\bibinfo{person}{Haoren Zhu}, \bibinfo{person}{Shih-Yang Liu}, \bibinfo{person}{Pengfei Zhao}, \bibinfo{person}{Yingying Chen}, {and} \bibinfo{person}{Dik~Lun Lee}.} \bibinfo{year}{2022}\natexlab{}.
\newblock \showarticletitle{Forecasting asset dependencies to reduce portfolio risk}. In \bibinfo{booktitle}{\emph{Proceedings of the AAAI Conference on Artificial Intelligence}}, Vol.~\bibinfo{volume}{36}. \bibinfo{pages}{4397--4404}.
\newblock


\end{thebibliography}

\end{document}